\documentclass[10pt,aps,prb,twocolumn,nisuperscriptaddress, nolongbibliography]{revtex4-2}

\usepackage{amsmath}

\usepackage{lmodern}
\usepackage[T1]{fontenc}
\setcitestyle{super}
\usepackage{soul}
\usepackage[utf8]{inputenc}
\usepackage{graphicx}
\usepackage{xcolor}
\usepackage{textgreek}
\usepackage{siunitx}

\usepackage[colorlinks=true,bookmarks=false,citecolor=blue,urlcolor=blue,linkcolor=red]{hyperref}

\begin{document}

\title{Silicon avalanche transition edge bolometer: approaching the thermodynamic limit for uncooled long-wave infrared detection}

\author{Wenjun Deng$^{1,2}$, Renwen Yu$^{3,4}$, Guanyu Han$^{1,2}$, Ziyao Feng$^{1,2}$, Yu Wang$^{1,2}$, Shanhui Fan$^3$, and Qiushi Guo${^{1,2,\dagger}}$\\
\vspace{3mm}
\textit{
$^1$Photonics Initiative, Advanced Science Research Center, City University of New York, NY, USA. \\
$^2$Physics Program, Graduate Center, City University of New York, New York, NY, USA.\\
$^3$Department of Electrical Engineering, Ginzton Laboratory, Stanford University, Stanford, CA, USA. \\
$^4$CNRS, Ecole Centrale Lyon, INSA Lyon, Université Claude Bernard Lyon 1, CPE Lyon, INL, UMR 5270, F-69134, Ecully, France\\
}
$^\dagger$Email:\href{mailto:qguo@gc.cuny.edu}{qguo@gc.cuny.edu}
}

\date{\today}

\begin{abstract}
Bolometers transduce incident electromagnetic radiation into measurable electrical signals via radiation-induced heating in thermo-resistive materials. They are uniquely capable of detecting low-energy photons without cryogenic cooling, and are widely deployed for uncooled long-wave infrared (LWIR) radiation detection and thermal imaging. Although their fundamental detection limit is set by the thermodynamic fluctuations, state-of-the-art uncooled bolometers still operate well above this limit due to the presence of Johnson noise, $1/f$ noise and noises from the readout circuits. Here, we address this challenge by introducing a new uncooled LWIR bolometer concept—the silicon avalanche transition edge (SATE) bolometer. Operating near the steep current transition edge associated with avalanche breakdown, the device exhibits an ultra-high, positive temperature coefficient of resistance (TCR) of 330 $\%$/K, greatly suppressing the impacts of other noise sources. Even without any thermal insulation structures, the SATE bolometer delivers a high room-temperature responsivity up to 160 mA/W for 9.5 μm radiation, a noise equivalent power of 370 pW/$\sqrt{\mathrm{Hz}}$, and a strong electro-thermal feedback enlarged bandwidth of 77 kHz—performance unattainable with conventional bolometer materials with a TCR of $-1\sim-3~\%$/K. Our work establishes a promising thermo-electric transduction mechanism toward high-sensitivity, high-speed room-temperature thermal imaging and infrared spectroscopy using CMOS technology.
\end{abstract}

\maketitle



According to Planck’s law, all objects near room temperature can emit thermal radiation,  a form of electromagnetic radiation predominantly in the 8–14~μm long-wave infrared (LWIR) spectral range\cite{howell2021thermal,xiao2021planck,xiao2020depth}. As such, LWIR detectors and imagers can enable vision even in complete darkness without the need for active illumination\cite{1033764,niklaus2008mems,bao2023heat}. However, room-temperature LWIR detection using photon detectors remains challenging because the photon energy is comparable to the energy scale of ambient thermal emission, causing the photo-generated carrier population to be readily overwhelmed by thermally excited carriers\cite{rogalski2005hgcdte,rogalski2000infrared}. In contrast, bolometers detect LWIR radiation by measuring the resulting change in the electrical property of a material induced by photo-thermal heating\cite{richards2004bolometers,mousa2026ultrabroadband,bauer2025exploiting, huang2026thermal}. This approach offers broad spectral coverage across the electromagnetic spectrum, particularly in regimes where photon detectors would otherwise require cryogenic cooling\cite{rogalski2026new,rogalski2000infrared}. Compared to optomechanical LWIR detectors\cite{das2023thermodynamically,vicarelli2022micromechanical,xi2025room,hui2016plasmonic,laurent201812,yi2013plasmonically} with optical or RF readout schemes, bolometers with simple current or voltage readout can be easily scaled into large-format focal plane arrays for thermal imaging\cite{kruse2001uncooled}. 
\begin{figure*}[ht]
\centering
\includegraphics[width=0.93\linewidth]{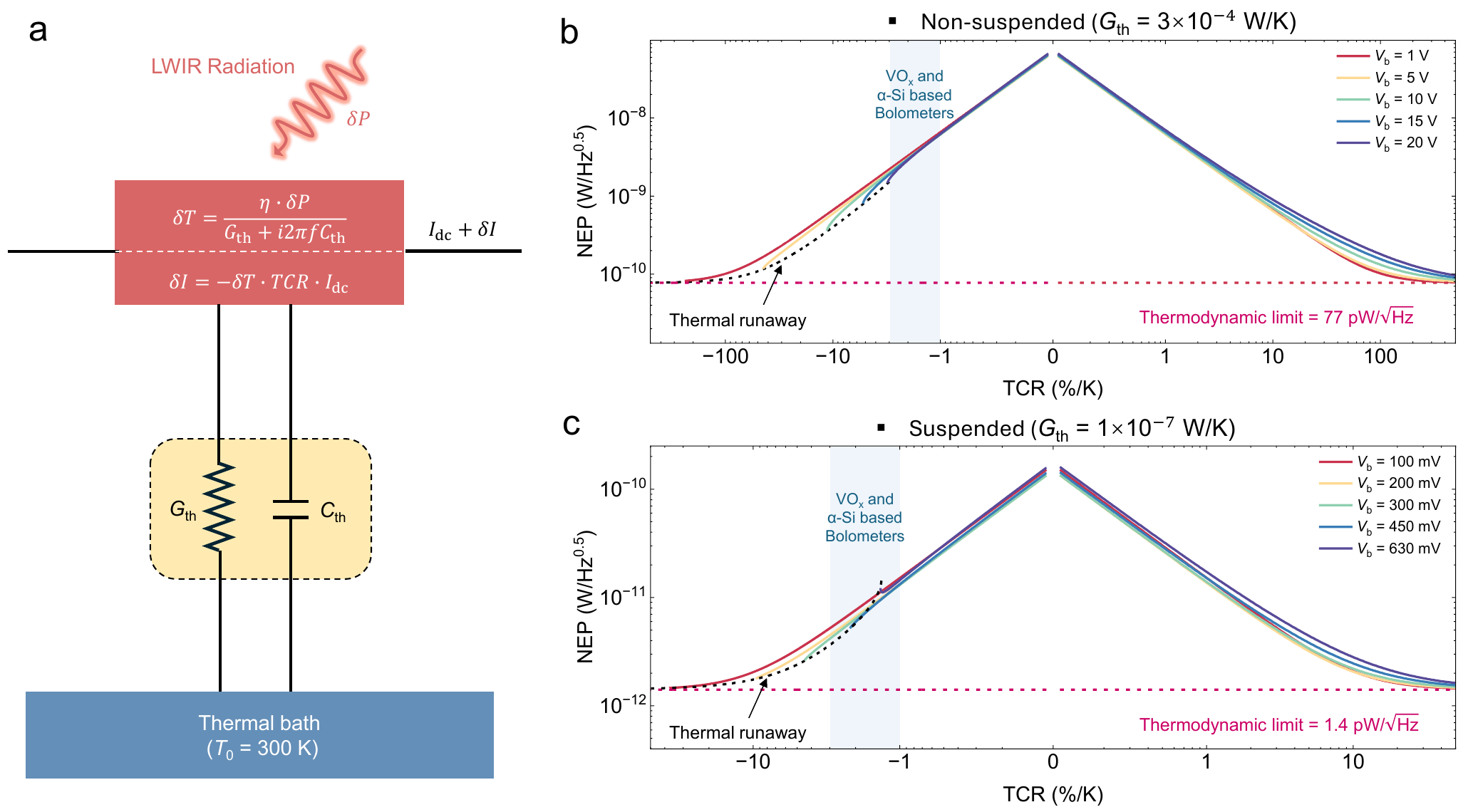}
\vspace{-5pt}
\caption{\textbf{Harnessing the ultra-high, positive TCR to approach the thermodynamic limit.} \textbf{a}, Physical model of a voltage-biased bolometer in response to LWIR radiation sinusoidally modulated at frequency $f$. \textbf{b}, Calculated NEP as a function of TCR at 1 kHz for non-suspended conventional bolometers ($G_\mathrm{th} = 3\times10^{-4}$ W/K) at various $V_\mathrm{b}$. \textbf{c,} Calculated NEP as a function of TCR at 30 Hz for suspended conventional bolometers ($G_\mathrm{th} = 1\times10^{-7}$ W/K) at various $V_\mathrm{b}$. In b and c, the thermodynamic limits are marked by the pink dashed lines. The operating regimes of VO$_\text{x}$ and $\alpha$-silicon bolometers are marked by the blue shadowed regions, corresponding to TCR values from $-1$ to $-3~\%$/K. Thermal runaway (black dashed lines) occurs when $V_\mathrm{b}$ exceeds certain threshold so that the ETF loop gain $\mathcal{L}=P_\mathrm{J}\cdot\mathrm{TCR} /G_\mathrm{th}$ is larger than 1. A higher |TCR| leads to thermal runaway at a lower $V_\mathrm{b}$.}
\vspace{-10pt}
\label{Fig1}
\end{figure*}

The ultimate detection limit of a bolometer is set by the thermal fluctuation noise, which arises from the stochastic exchange of thermal energy between the thermo-resistive material and the thermal bath at thermal equilibrium\cite{callen1951irreversibility,stewart2021nanophotonic}. This establishes a thermodynamically limited noise equivalent power (NEP) given by $\mathrm{NEP}_\mathrm{TF}=\sqrt{4 k_{\mathrm{B}} T^2 G_{\mathrm{th}}}$, which depends only on the device temperature $T$ and the thermal conductance $G_{\mathrm{th}}$ between the thermo-resistive material and the thermal bath\cite{mather1982bolometer,watts2007microphotonic}. Yet, the state-of-the-art uncooled bolometers based on thermo-resistive materials, such as vanadium oxide (VO$_x$) and amorphous silicon ($\alpha$-Si), operate well above this thermodynamic limit. This is because their relatively small ($\sim -1$ to $-3~\%$/K) temperature coefficient of resistance (TCR)\cite{rogalski2026new,yadav2022advancements,niklaus2008performance} limits the transduction of photo-induced heating into electrical resistance changes, causing Johnson noise and low-frequency $1/f$ noise\cite{niklaus2008performance} to dominate over thermal fluctuation noise.

Here, we first show that the NEP of uncooled bolometers can be reduced toward the thermodynamic limit by substantially enlarging the magnitude of TCR, which serves as a novel way to optimize the NEP. We further reveal that the sign of TCR plays a critical role in the NEP management, and a positive TCR is more favorable for voltage-biased bolometers. Building on this insight, we break the material-limited TCR constraint by exploiting the avalanche current transition-edge effect in a silicon-on-insulator (SOI) platform, achieving an ultra-high, positive TCR of $\sim330~\%/\mathrm{K}$. Such a highly efficient thermo-electrical transduction enables the bolometer to operate in a regime where shot noise and thermal fluctuation noise dominate over other noises. Even without any thermal isolation, our device achieves a room-temperature NEP of $370~\mathrm{pW}/\sqrt{\mathrm{Hz}}$ that is only $\sim4.8\times$ above the thermodynamic limit of $77~\mathrm{pW}/\sqrt{\mathrm{Hz}}$. Moreover, the large positive TCR provides a strong negative electro-thermal feedback that simultaneously enhances device stability and extends the operational bandwidth.

\noindent\textbf{Dependence of NEP on TCR.} To quantify how various noise mechanisms constrain the detection limit, we start from a generic physical model of a voltage-biased bolometer. As shown in Fig.~\ref{Fig1}a, a sinusoidally modulated LWIR radiation with power of $\delta P$ and a modulation frequency $f$ is first absorbed by the active region that has an absorption efficiency of $\eta$ ($\%$). Upon the absorption, the active region experiences a temperature perturbation $\delta T=\eta \cdot \delta P/(G_{\mathrm{th}}+i2\pi f C_{\mathrm{th}})$, where $C_{\mathrm{th}}$ is the thermal capacitance of the device, $G_{\mathrm{th}}$ is the thermal conductance between the device and the thermal bath (assumed to be at a temperature $T_0$=300 K). This temperature perturbation further affects the DC-resistance ($R_\mathrm{dc}$) of the thermo-resistive material via its TCR, defined as $\mathrm{TCR}= R_\mathrm{dc}^{-1}\cdot \partial R_\mathrm{dc}/\partial T$ (\%/K). This leads to a photocurrent of $\delta I=-I_\mathrm{dc}\cdot \mathrm{TCR}\cdot \delta T$ at a fixed bias voltage, and an extrinsic LWIR responsivity of $\mathcal{R}_\mathrm{ext} =\delta I/\delta P$. Considering all noise sources and the electro-thermal feedback (ETF) due to the device's Joule heating \cite{mather1982bolometer,irwin1995application,govenius2016detection}, the NEP can be calculated as:
\begin{equation}
\small
\mathrm{NEP}=\frac{|\alpha G_{\mathrm{th}}^2|}{|\eta\widetilde{G}|}\sqrt{\frac{|\alpha|^2e^{\beta} K_{\mathrm{f}}}{\text{TCR}^2f}+\frac{4k_{\mathrm{B}}T_0\left|\alpha-\beta\right|^2}{\beta\cdot\text{TCR}\cdot G_{\mathrm{th}}}+\frac{4k_{\mathrm{B}}T_0^2}{G_{\mathrm{th}}}},
\label{Eq1}
\end{equation}
where the first, second, and third terms represent the contributions from the $1/f$ noise\cite{Hooge1994}, Johnson noise\cite{nyquist1928thermal}, and thermal fluctuation noise, respectively. $\alpha={\widetilde{G}(P_\text{J}+G_{\mathrm{th}}T_0)/(G_{\mathrm{th}}^2T_0)}$, $\beta=P_\text{J}\cdot\text{TCR}\cdot T_0/(P_\text{J}+G_\text{th}T_0)$, $\widetilde{G}=G_\mathrm{th}+i2\pi fC_\mathrm{th}$, $K_\mathrm{f}=1.047\times10^{-11}$ is the flicker noise coefficient of the $1/f$ noise\cite{hooge20021}, $k_{\mathrm{B}}$ is the Boltzmann constant, and $T_0$ is the thermal bath temperature\cite{mather1982bolometer}. The bias voltage $V_\mathrm{b}$ determines the Joule heating power $P_{\mathrm{J}}$. 

\begin{figure*}[ht]
\vspace{-5 pt}
\centering
\includegraphics[width=0.96\linewidth]{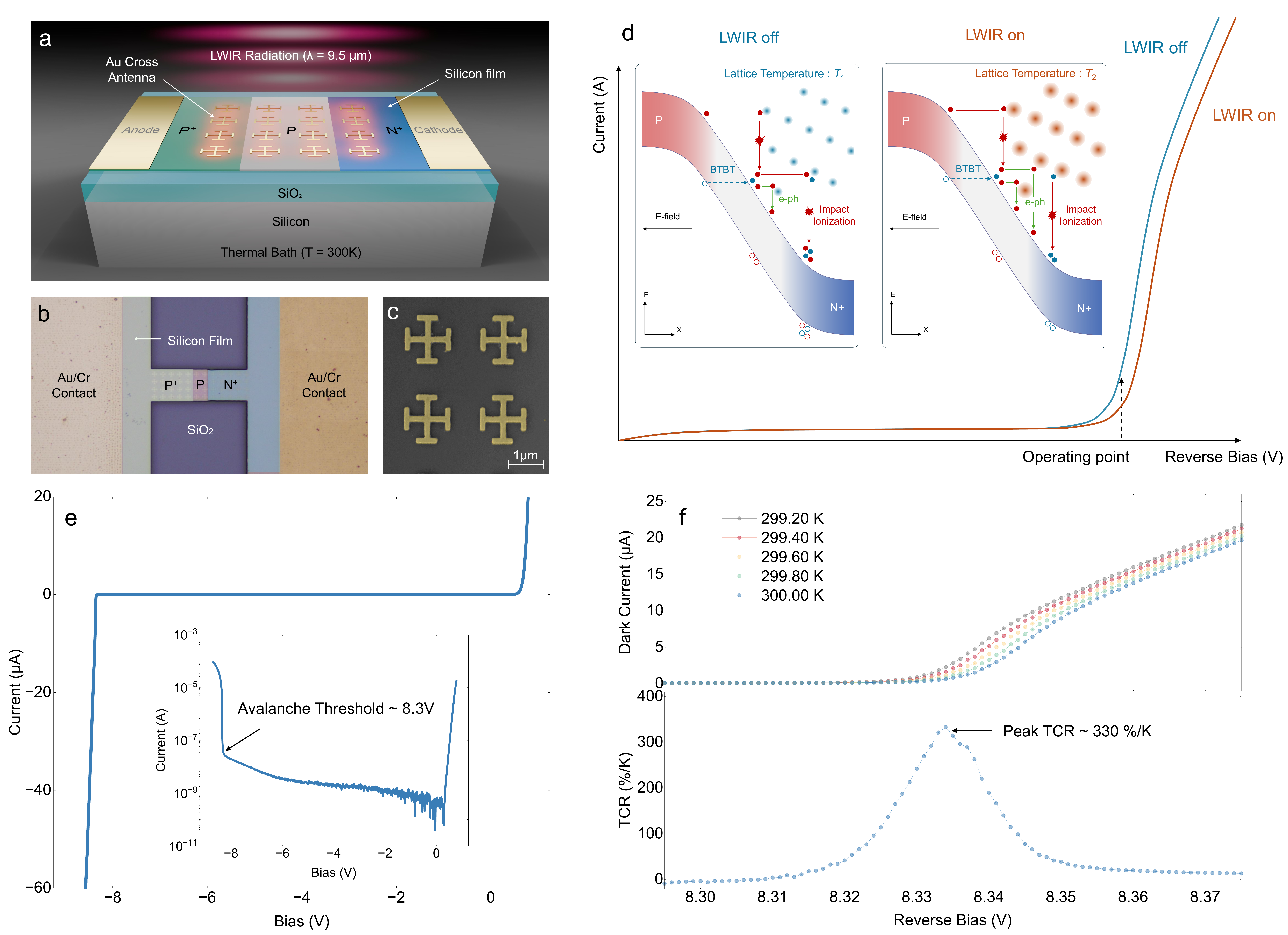}
\vspace{-6 pt}
\caption{\textbf{SATE bolometer with ultra-high TCR.} \textbf{a,} Schematic of a SATE bolometer pixel. A laterally doped  $p^+\text{-}p^{~}\text{-}n^+$ diode fabricated on the SOI is reversely biased via two Au/Cr electrodes to sense the LWIR radiation with enhanced absorption by the Au plasmonic antenna array. \textbf{b,} False colored optical microscope image of a single SATE bolometer pixel. \textbf{c,} False-colored scanning electron microscope (SEM) image of gold antennas (yellow). \textbf{d,} Illustration of TCR enhancement by avalanche transition edge effect. Inset: Minority carriers undergo impact-ionization-induced multiplication in the high-electric-field intrinsic region, resulting in a pronounced difference in current with and without LWIR-induced temperature rise near the avalanche breakdown threshold. \textbf{e,}  Measured I–V characteristics of the SATE bolometer under the dark condition. Inset: I–V characteristics in log scale. \textbf{f,} Upper panel: Temperature-dependent I–V characteristics measured with  0.20 K intervals. Lower panel: extracted TCR as a function of reverse bias voltage.} 
\label{Fig2}
\vspace{-10pt}
\end{figure*}

Figure~\ref{Fig1}b and c visulize the calculated NEP as a function of TCR at different $V_\mathrm{b}$. Here, we assumed that the room temperature bolometer resistance remains 100 k$\Omega$ as the TCR is varied. In the negative-TCR regime, when |TCR| is small, the bolometer is primarily limited by 1/$f$ and Johnson noises. As |TCR| increases, the contributions of these noise sources are reduced. As shown in the blue-shadowed regions, given the limited TCR of $-1$ to $-3~\%$/K, the NEP of conventional VO$_\text{x}$ and $\alpha$-$\mathrm{Si}$ bolometers are at least 25 and 5 times higher than the corresponding thermodynamic limits for the non-suspended ($G_{\mathrm{th}} = 3 \times 10^{-4}$ W/K) and suspended ($G_{\mathrm{th}} = 1 \times 10^{-7}$ W/K) cases, respectively. Moreover, conventional bolometers with a negative TCR are susceptible to thermal runaway at higher $V_\mathrm{b}$\cite{brandao2001stability}. This is because a positive ETF loop can be established: a temperature increase reduces the device resistance, resulting in more Joule heating, which further raises the temperature and destabilizes the operating point. Details regarding the calculation and analysis of Fig.~\ref{Fig1}b and c are in Supplementary Information Section~1. 
\vspace{0.5mm}

Clearly, by comparing Fig.~\ref{Fig1}b and c, a common route toward approaching the thermodynamic limit is to reduce $G_{\mathrm{th}}$ through aggressive thermal-isolation structures, as widely employed in state-of-the-art mcirobolometers. However, this inevitably increases the thermal time constant ($\tau_\mathrm{th} = C_{\mathrm{th}} / G_{\mathrm{th}}$) to the millisecond scale\cite{richards2004bolometers,gitelman2009cmos,kohin2004performance}, which limits the operating bandwidth below $100\,\mathrm{Hz}$ and restricts applications such as tracking fast-moving objects \cite{richards2005applications,Lee2022MotionBlurThermalDetector} or high-throughput analysis\cite{hefner2001high}. As shown in the right half of Fig.~\ref{Fig1}b and c, a promising route emerges if a device can achieve an ultra-high, and positive TCR without significantly increasing its resistance. Such a combination enables the bolometer to approach the thermodynamic limit without sacrificing speed. Moreover, a positive TCR naturally leads to a negative ETF loop that stabilizes the operating point.

\begin{figure*}[ht]
\vspace{-5 pt}
\centering
\includegraphics[width=1\linewidth]{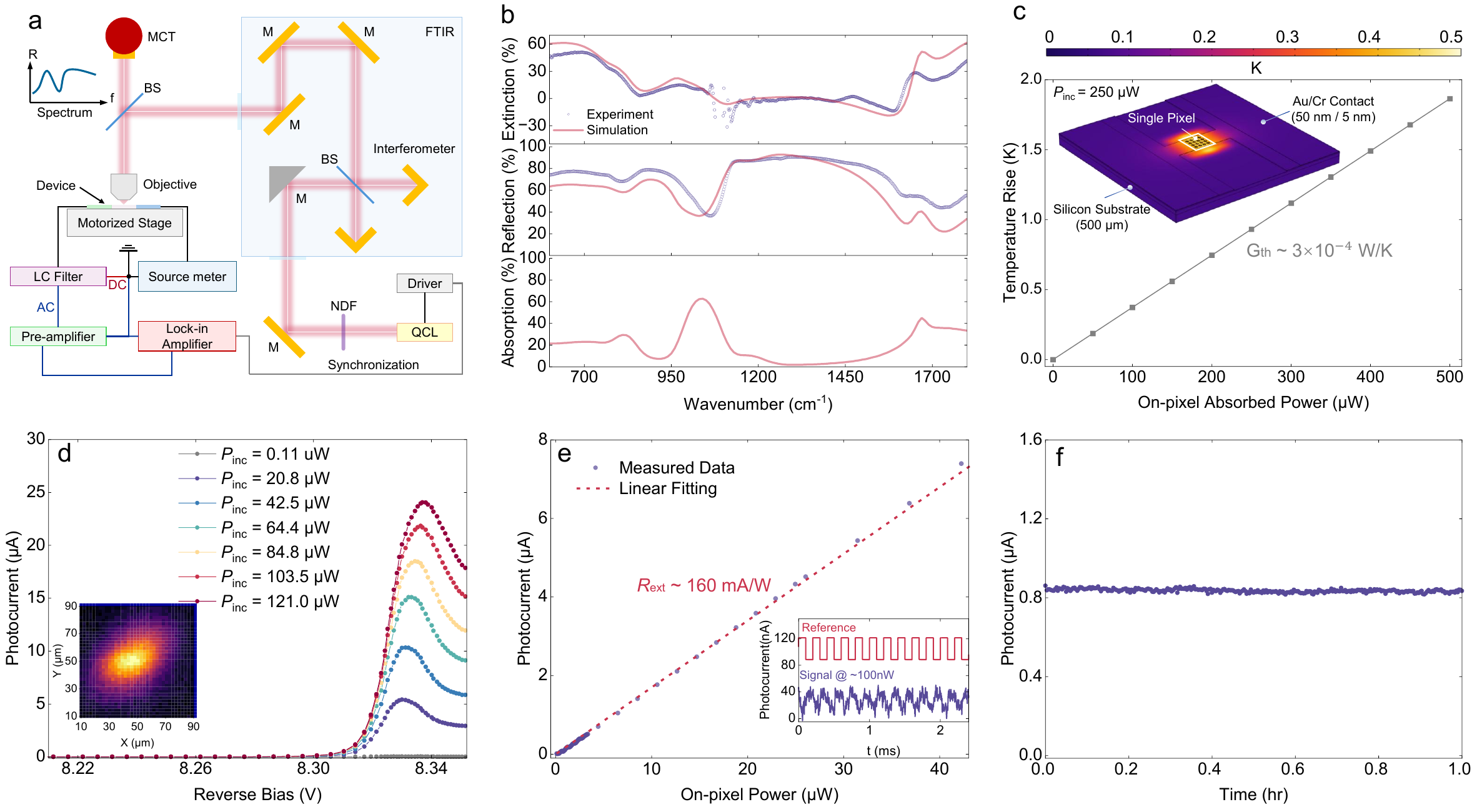}
\vspace{-15 pt}
\caption{\textbf{9.5 μm LWIR photoresponse.} \textbf{a}, Schematic of the measurement setup. The QCL beam is focused onto the SATE bolometer pixel via the beam path of the FTIR. The QCL beam is AC-modulated at $f=5~\mathrm{kHz}$ via the laser driver. The SATE bolometer is voltage-biased by a source meter. The AC part of photocurrent was picked out through an LC filter, amplified by the transimpedance amplifier (TIA), and read out by the lock-in amplifier in synchronization with the QCL current driver. M: gold mirror. BS: beam splitter. NDF: neutral density filter. MCT: liquid nitrogen cooled HgCdTe detector. \textbf{b}, Optical spectra of the plasmonic antenna array measured by FTIR (blue symbols), together with the corresponding simulation results (red solid line). \textbf{c}, Simulated temperature rise as a function of on-pixel absorbed LWIR power. An effective $G_\mathrm{th}=3 \times 10^{-4}$ W/K  is extracted  from the linear fit (gray solid line). Inset: simulated 3-D distribution of the temperature rise in response to 250~μW of on-pixel incident power.  \textbf{d}, Measured photocurrent as a function of reverse bias voltages under different on-pixel incident powers. Inset: Profile of QCL beam spot measured by scanning the SATE bolometer pixel. \textbf{e}, Power-dependent photocurrent measured at $V_\mathrm{b} = -8.34$~V. A responsivity of $\mathcal{R}_\mathrm{ext}=160$ mA/W is extracted from the linear fit (red dashed line). Inset: measured photocurrent waveform at an on-pixel power of 100 nW. \textbf{f}, Photocurrent stability test over 1 hour at $V_\mathrm{b} = -8.325$~V.} \label{Fig3}
\vspace{-10pt}
\end{figure*}

\noindent\textbf{Silicon avalanche transition edge (SATE) bolometers with ultra-high positive TCR.} In practice, the sign and the achievable TCR value are usually constrained by the intrinsic properties of materials, such as the thermal activation energy, the band gap \cite{chen2026bias,niklaus2008mems}, and the weak dependence of carrier mobility on temperature\cite{sze2021physics,ziman1979principles,stewart2021nanophotonic}. To break such a material-constrained TCR limit, one promising approach consists of exploiting the semiconductor device physics\cite{chen2026bias}. For instance, it is known that the avalanche breakdown voltage threshold is highly sensitive to temperature\cite{crowell1966temperature,tan2010temperature}, suggesting that a sharp current transition edge effect could serve as an efficient thermo-electric transduction mechanism with ultra-high TCR. As illustrated in Fig.~\ref{Fig2}a, our SATE bolometer device is composed of a lateral $p^+\text{-}p^{~}\text{-}n^+$ junction fabricated on an SOI wafer, with metal contacts on the $p^+$ and $n^+$ regions. A gold (Au) cross-shaped plasmonic antenna array is integrated on top of the  $p^+\text{-}p^{~}\text{-}n^+$ junction to facilitate the absorption of LWIR radiations. An optical microscope image of a fabricated SATE bolometer pixel is shown in Fig.~\ref{Fig2}b. We patterned the bolometer pixel size to be 12 µm $\times$12 µm in order to match the pitch of the state-of-the-art microbolometer pixels\cite{teledyneflir2025bosonplus,lynred2022atto640d}. We used a spin-on-dopant (SOD) doping process\cite{jaeger2002microelectronic}  to form the lateral  $p^+\text{-}p^{~}\text{-}n^+$ junctions on the device layer, with doping concentrations of $3\times10^{18}~\mathrm{cm}^{-3}$, $3.1\times10^{17}~\mathrm{cm}^{-3}$, and $5\times10^{18}~\mathrm{cm}^{-3}$, respectively (see Methods for details). The width of the $p$ region is 5 µm. On top of the silicon layer, the Au cross antennas with C4 rotational symmetry are fabricated in a square $4\times4$ lattice with a periodicity of 2.5 µm (Fig.~\ref{Fig2}c), allowing for polarization-independent absorption enhancement of LWIR radiation at 9.5 µm.

The operating principle of the SATE bolometer is summarized in Fig.~\ref{Fig2}d inset. In a reverse-biased $p-n^+$ junction, minority carriers, either injected from the $p$ and $n^+$ sides or generated through band-to-band tunneling (BTBT, blue dashed arrows)~\cite{kane1960zener}, were accelerated by electric field within a carrier mean free path. Carriers gaining kinetic energy exceeding the bandgap can ionize lattice atoms through collisions, a process known as impact ionization~\cite{grant1973ionization}(red arrows with sparks). In contrast, carriers with insufficient energy are scattered by the lattice through electron-phonon scattering (e-ph, green-solid arrows). When the reverse voltage bias reaches a certain threshold, the newly generated carriers from impact ionization can repeatedly ionize lattice atoms, leading to a dramatic exponential increase in the carrier population across the depletion region and avalanche breakdown. During this multiplication process, the LWIR-radiation-induced lattice temperature rise increases the probability of electron-phonon scattering~\cite{crowell1966temperature}, thereby reducing the population of carriers participating in impact ionization. As a result, near the avalanche current transition edge, a small increase of device temperature can lead to a large current drop (Fig.~\ref{Fig2}d), thereby enabling an ultra-high and positive TCR.

Figure~\ref{Fig2}e shows the measured the current-voltage (I-V) characteristics of a SATE bolometer pixel under dark condition, which exhibits a $-8.3$~$\mathrm{V}$ avalanche breakdown threshold. We also measured the temperature-dependent I-V characteristics under dark conditions near the avalanche breakdown from 299.2 K to 300 K, with a temperature interval of 0.20 K controlled by a temperature controller underneath the device. As shown in Fig.~\ref{Fig2}f upper panel, the avalanche breakdown voltage threshold clearly increases with temperature by $\sim$ 8 mV/K. Figure~\ref{Fig2}f lower panel shows that an ultra-high peak TCR of 330 \%/K is obtained at -8.325 V, which is $\sim$ 100 times higher in magnitude than that of VO$_\text{x}$ and $\alpha$-$\mathrm{Si}$. It is worth noting that conventional thermo-resistive materials typically exhibit a negative TCR due to their thermally activated carrier transport\cite{abdel2019temperature,czerwinski2003activation,chen2026bias}. In contrast, our SATE bolometer exhibits a positive TCR that uniquely enables strong negative ETF, as will be discussed below.

\begin{figure*}[ht]
\vspace{-5 pt}
\centering
\includegraphics[width=0.98\linewidth]{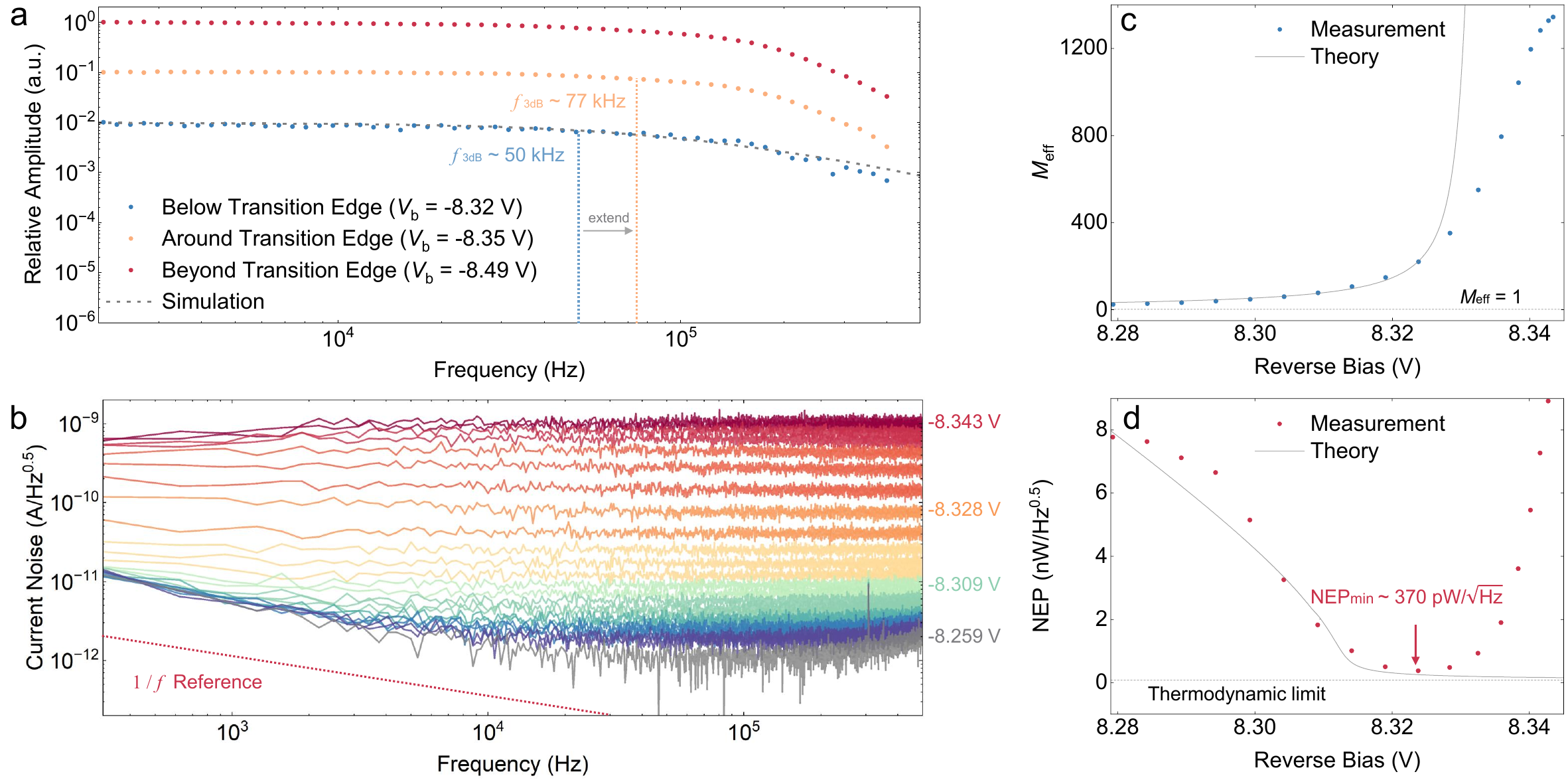}
\vspace{-12 pt}
\caption{\textbf{Frequency response and noise characteristics.} \textbf{a,} Simulated (grey solid line) and measured (colored symbols) frequency responses below ($V_\mathrm{b} = -8.32$~ V), around ($V_\mathrm{b} = -8.35$~V), and above ($V_\mathrm{b} = -8.49$~V) the avalanche transition edge. The photocurrent amplitudes are vertically offset to facilitate comparison.  \textbf{b,} Dark current noise spectral density measured at different reverse bias voltages. The red dashed line marks the $1/f$ trend. 
\textbf{c,} Symbols: extrapolated effective excess-noise $M$ factor ($M_\text{eff}$) as a function of reverse bias voltage. Solid line: theoretical $M_\text{eff}$ of an ideal step p-n junction. \textbf{d,} Symbols: measured NEP at 1 kHz. The minimum measured NEP is 370 pW/$\sqrt{\mathrm{Hz}}$ at $V_\mathrm{b} = -8.324$~V. Solid line: theoretical NEP model calculated by Eq. (2).} 
\label{Fig4}
\vspace{-10pt}
\end{figure*}

\noindent\textbf{9.5 μm LWIR photoresponse}. Next, we characterize the LWIR photoresponse of the SATE bolometer by using the experimental setup shown in Fig.~\ref{Fig3}a. The Fourier-transform infrared spectrometer (FTIR) together with its integrated blackbody source, enables characterization of the device’s LWIR spectral response. Moreover, to measure the LWIR-induced photocurrent and responsivity, an external 9.5 µm quantum cascade laser (QCL) was guided into the FTIR and focused onto the device through the infrared objective. As shown in Fig.~\ref{Fig3}b, both the measured extinction spectrum (blue symbols) ($1-\mathcal{T}/\mathcal{T}_0$), and the reflection spectrum of the SATE bolometer pixel show good agreement with numerical simulations (red solid lines). Therefore, the device's absorption spectrum can be reasonably inferred using simulated absorption spectrum, which indicates an absorption of $\sim50\%$ at $9.5~\text{μm}$. Furthermore, we used COMSOL Multiphysics to simulate the temperature rise ($\Delta T$) in the pixel region induced by the incident LWIR radiation (see more details in Methods), taking into account both the optical absorption and the 3D heat transfer. As shown in Fig.~\ref{Fig3}c, the $\Delta T$ increases linearly with the absorbed on-pixel absorbed optical power. From the linear fit, we extrapolate an effective thermal conductance ($G_\mathrm{th}$) of $3 \times 10^{-4}$ W/K between the pixel region and the thermal bath. The $\Delta T$ distribution at thermal equilibrium is shown in the inset of Fig.~\ref{Fig3}c.

Next, we characterize the extrinsic responsivity ($\mathcal{R}_\mathrm{ext}$) of the SATE bolometer at 9.5 μm. To accurately determine the on-pixel incident power ($P_\mathrm{inc}$), we first mapped the beam intensity profile by mechanically scanning the bolometer pixel across the beam spot with a step size of 2 μm, while recording the photocurrent at each position. Using this approach, we obtained an elliptical-shaped beam profile of the 9.5 μm QCL shown in the inset of Fig.~\ref{Fig3}d. By combining the measured beam-spot intensity profile and the calibrated optical power, the on-pixel power at the center of the beam spot can be precisely calculated (see Methods for details). Figure~\ref{Fig3}d shows the measured photocurrent as a function of reverse bias voltage and $P_\mathrm{inc}$. Clearly, all curves exhibit negligible photocurrent when the reverse bias voltage is below 8.3 V, followed by a pronounced photocurrent peak around 8.33 V. This behavior is consistent with the avalanche-transition-edge-enhanced responsivity discussed above. The peak photocurrent occurs at a reverse bias near peak TCR, since the photocurrent is proportional to $\mathrm{TCR}\cdot I_\mathrm{dc}$. Figure~\ref{Fig3}e shows the measured photocurrent at $V_\mathrm{b}=-8.34$ V as a function of ${P_\mathrm{inc}}$, which exhibits a linear behavior when $P_\mathrm{inc}$ is varied from 100 nW to 40 μW. The linear fit yields a peak $\mathcal{R}_\mathrm{ext}$ of 160 mA/W. At the peak TCR, the steady-state temperature rise induced by Joule heating is estimated to be 0.11 K, according to $\Delta T=P_\mathrm{J}/G_\mathrm{th}$.

Notably, as shown by the inset of Fig.~\ref{Fig3}e, when biased near the peak TCR, the SATE bolometer can clearly resolve the photocurrent waveform generated by a modulated 9.5 μm laser beam with a ${P_\mathrm{inc}}$ as low as 100 nW. Moreover, since our device operates near the sharp avalanche transition edge, a natural question arises regarding the stability of the operating point and the corresponding photocurrent response. As shown in Fig.~\ref{Fig3}f, when biased at the peak TCR point, our device exhibits stable photoresponse for at least 1 hour with negligible photocurrent drift. Such a stability originates from the strong negative ETF enabled by the unusually large positive TCR.


\noindent\textbf{Frequency response and noise characteristics.} Here, we first measure the frequency response of photocurrent at different $V_\mathrm{b}$ by modulating the driving current of the QCL. In Fig.~\ref{Fig4}a, we plot three representative frequency responses measured below ($V_\mathrm{b}=-8.32~\mathrm{V}$), around ($V_\mathrm{b}=-8.35~\mathrm{V}$), and far above ($V_\mathrm{b}=-8.49~\mathrm{V}$) the avalanche transition edge. Below the transition edge, the measured frequency response (red symbols) shows good agreement with the simulated result (black solid line) obtained using COMSOL (see more details in Supplementary Section~2.2). They both exhibit an intrinsic 3-dB cutoff frequency of $\sim$ 50 kHz, which is primarily limited by the large thermal capacitance of the 2-\textmu m-thick $\mathrm{SiO_2}$ layer beneath the silicon device layer, where most of the optical absorption for the 9.5 $\mu$m incident light and the resulting temperature rise occur. The faster measured roll-off compared with the simulation at higher frequencies is attributed to the 400~kHz cutoff frequency of the laser driver.  Intriguingly, both the frequency responses measured around and above the transition edge (blue symbols) show a significantly extended 3-dB cutoff frequency of 77 kHz, corresponding to a thermal response time $\tau_\mathrm{th}$ of 2.07 μs. This behavior arises from the negative ETF enabled by the large positive TCR of our SATE bolometer, which enhances the operating bandwidth by a factor of ($1+P_\mathrm{J}\cdot\mathrm{TCR} /G_\mathrm{th}$) (see Supplementary Section 1.2.1 for more details).

Unlike conventional bolometers based on thermo-resistive materials, our SATE bolometers involve semiconductor $p\text{-}n$ junctions, which leads to additional current shot noises due to the discrete and stochastic transport of charge across the junction\cite{vanderziel1955shotnoise}. Moreover, due to the carrier multiplication process near the avalanche transition edge, the excess shot noise spectral density can be expressed as $\sqrt{2qM^2F(M)I_\text{0}}$\cite{mcintyre1966multiplication}, where $q$ is the elementary charge, $I_0$ is the reverse-biased current without considering impact ionization, and $M$ is the multiplication factor due to impact ionization gain. Since $MI_0$ equals to the measured DC current $I_\text{dc}$ and $F(M)$ depends on $M$, we rewrite the excess shot noise spectral density as $\sqrt{2qM^2_\text{eff}I_\mathrm{dc}}$, where $M_\text{eff}=\sqrt{MF(M)}$. To quantify the total dark current noise spectral density and the NEP of our device, we employed a custom-built noise measurement setup fully calibrated against the Johnson noise of carbon resistors, as detailed in Supplementary Section 3. Figure~\ref{Fig4}b shows the measured current noise amplitude spectral density at several $V_\mathrm{b}$ from $-8.259$ to $-8.343~\mathrm{V}$ across the avalanche transition edge. Below the breakdown threshold, the spectrum features a pronounced low-frequency $1/f$ component and a nearly flat (white) frequency response at higher frequencies. When the device is reverse-biased above $8.3~\mathrm{V}$, the noise spectrum is dominated by the frequency-independent excess shot noise.

By fitting the measured current noise spectral density with a model incorporating $1/f$ noise, shot noise, Johnson noise, and thermal fluctuation noise (see Supplementary Section~3.3), we extracted $M_\text{eff}$ as a function of reverse bias voltage, as shown in Fig.~\ref{Fig4}c (symbols). The solid line in Fig.~\ref{Fig4}c gives the theoretical calculation of $M_\text{eff}$ derived from the impact ionization integral for an ideal step $p$-$n$ junction (see Supplementary Section~3.4). According to both the measured voltage-dependent current noise spectral density ($\delta I_\mathrm{n}$, in unit of $\mathrm{A}/\sqrt{\mathrm{Hz}}$) and the voltage-dependent $\mathcal{R}_\mathrm{ext}$ (in unit of A/W) obtained from Fig.~\ref{Fig3}d, the NEP can be calculated as a function of the reverse bias voltage as $\text{NEP}=\delta I_\mathrm{n}/\mathcal{R}_\mathrm{ext}$. As shown in Fig.~\ref{Fig4}d, at 1 kHz, a minimum measured NEP of \(370~\mathrm{pW/\sqrt{Hz}}\) is obtained at $V_\mathrm{b}=-8.324~\mathrm{V}$, with a TCR $\sim$ 100~$\%$/K and $M_\text{eff}\sim$ 200. Notably, the minimum NEP we measured is only $\sim$ 4.8 times higher than the thermodynamic limit of \(77~\mathrm{pW/\sqrt{Hz}}\), calculated at \(T = 300~\mathrm{K}\) and \(G_{\mathrm{th}} = 3 \times 10^{-4}~\mathrm{W/K}\). Around the transition edge, the measured NEP also agrees well with the theoretical estimation (solid line in Fig.~\ref{Fig4}d), which is calculated as (see Supplementary Section~3.4 for more details): 
\begin{equation}
\footnotesize
\mathrm{NEP}
=
\frac{1}{\eta}\sqrt{
\frac{
2qI_\mathrm{dc}M_\text{eff}^2(\frac{\partial V}{\partial I_\mathrm{dc}})^2\left|\widetilde{G}\right|^2
}
{
\left|\partial V/{\partial T}\right|^2
}
+\frac{4k_\text{B}T\frac{\partial V}{\partial I_\mathrm{dc}}\left|\widetilde{G}\right|^2}
{\left|\partial V/\partial T\right|^2}
+4k_\text{B}T^2G_\text{th}},
\label{Eq2}
\end{equation}
where $V$ is the voltage across the junction. The first, second, and the third term are contributed by excess shot noise, Johnson noise, and thermal fluctuation noise, respectively, where the excess noise is dominant over the rest. We attribute the increase in the measured NEP above 8.33 V in Fig.~\ref{Fig4}d to the device’s stronger nonlinear response. In this regime, the first-order small-signal approximation used to derive Eq.~\ref{Eq2} is no longer valid.
\vspace{1.0mm}

Finally, we also would like comment on the NEP at even lower frequencies below 1 kHz where $1/f$ noise becomes important. From the measured current noise spectral density at $I_\mathrm{dc}$~$\sim66~\mathrm{nA}$ (bottom gray line in Fig.~\ref{Fig4}b, corresponding to the $V_\mathrm{b}$ of $-8.259$ V), we extracted the flicker noise coefficient $K_\mathrm{f}$ as $\sim$ 5$\times10^{-6}$ (see Supplementary Section~3.3). According to Hooge's empirical relation, $K_\mathrm{f}$ is inversely proportional to the total number of carriers in the active region\cite{hooge1969oneoverf}. Around the avalanche transition edge, the total number of carriers is proportional to the DC current. Therefore, $K_\mathrm{f}$ at other operation currents can be obtained by multiplying the $K_\text{f}$ we extracted at $I_\mathrm{dc}$~$\sim66~\mathrm{nA}$ by a current ratio $66 [\text{nA}]/I_\text{dc}$. At the minimum NEP operating point ($I_\mathrm{dc}$~$\sim111~\mathrm{nA}$), the  $K_\mathrm{f}$ is estimated to be \(2.97\times10^{-6}\). This translates to current noise spectral density of \(194~\mathrm{pA/\sqrt{Hz}}\) and an NEP of \(1.8~\mathrm{nW/\sqrt{Hz}}\) at 1 Hz. As a comparison, conventional thermo-resitive materials with TCR from $-1$ to $-3~\%$/K would exhibit a minimum NEP of \(96.8~\mathrm{nW/\sqrt{Hz}}\) at 1 Hz without the use of mechanical suspension structures.
\vspace{1.0mm}

\noindent\textbf{Discussion and outlook.} In this work, instead of relying on conventional material engineering approaches, we exploit avalanche transition edge effect in silicon to realize an ultra-efficient thermo-electrical transduction mechanism, with an exceptionally high TCR up to 330 $\%$/K. This leads to a large current change in response to small temperature perturbation induced by the LWIR radiation, thus substantially enhancing the LWIR responsivity and allowing shot noise and thermal fluctuation noise to dominate over other noise sources.  As a result, the power detection limit can get close to the thermodynamic limit even in absence of any mechanically suspended thermal isolation structures. Notably, this mechanism is in stark contrast to simply enhancing the responsivity using an external amplifier, which simultaneously amplifies all noises.
\vspace{0.6mm}

We also want to highlight some other benefits uniquely enabled by the ultra-high, positive TCR of our SATE bolometer in the nonlinear avalanche transition regime. First, unlike the conventional expectation that devices operating near a sharp current transition edge may suffer from instability, our device exhibits negligible variation in photoresponse over one hour (Fig. \ref{Fig3}f). Such an excellent stability originates from the strong self-stabilizing negative ETF mechanism enabled by the large and positive TCR of our device: a sudden increase in device current leads to more Joule heating and increase in resistance, which in turn suppresses the current flow. Second, the strong negative ETF offers an unexplored revenue toward overcoming the intrinsic bandwidth limitation of conventional uncooled bolometers, as experimentally validated by the results shown in Fig.~\ref{Fig4}a. 
\vspace{1mm}

Our current device has already achieved the state-of-the art figure of merit (NEP$\cdot \tau_\mathrm{th}$)\cite{laurent201812,shao2026nonlinear,skidmore2014uncooled} of $7.7\times10^{-16}$ W$\cdot$Hz$^{-3/2}$ for uncooled bolometers.  Looking forward, by combining our avalanche transition-edge bolometer concept with silicon-based thermal isolation structures\cite{chen2021ultrafast}, optimized LWIR plasmonic absorbers\cite{audhkhasi2025compact}, and a thinner silicon layer with smaller thermal capacitance, it is possible to overcome the sensitivity–speed trade-off that have long constrained conventional microbolometers\cite{laurent201812}. For instance, assuming a positive TCR of 100$\%$/K near the avalanche transition edge with $M_\text{eff}=200$ and a very modest thermal insulation with $G_\mathrm{th}=3\times10^{-6}$ W/K,  it is sufficient to reach a thermodynamics-limited NEP of 2.53 pW/$\sqrt{\mathrm{Hz}}$, which is on par with the commercially available microbolometers with an aggressive thermal insulation of $G_\mathrm{th}=5\times10^{-8}$ W/K\cite{fisette2018customized}. Simultaneously, a stronger negative ETF due to suspension can lead to significantly larger modulation bandwidth compared to existing microbolometers. We believe that our CMOS-compatible SATE bolometer architecture could have broad implications for future high-speed thermal imaging, fast target tracking\cite{richards2005applications}, and cost-effective mid-infrared spectroscopy\cite{jeon2021development}.

\section*{Methods}
\noindent \textbf{Device fabrication.} The SATE bolometer was fabricated on a silicon-on-insulator (SOI) wafer with a 600~nm device layer (Resistivity: 10 $\Omega$/cm), a 2~μm buried oxide layer, and a 500~μm silicon substrate. The device layer was first lightly p-doped by spin-coating a spin-on dopant solution (B153, Filmtronics) diluted 15× in isopropyl alcohol (IPA), followed by thermal activation at $900~^{\circ}\mathrm{C}$ in an Ar/O$_2$ (9:1) ambient for 10 minutes. Heavily doped $p^+$ and $n^+$ regions were then formed using un-diluted spin-on dopant solution (B153, P509 Filmtronics) followed by the same thermal activation process, with HSQ serving as a diffusion barrier defined by 50~keV electron-beam lithography system (Elionix ELS-G50). After doping each region, the HSQ mask was removed by dipping the sample in buffered oxide etchant (BOE) for 20 minutes. We measure the doping concentration of each region by using a four-point probe sheet resistance tester (Huanyu). Then, we pattern the isolation window of each device by electron-beam lithography using PMMA (A6) as the e-beam resist. After developing in MIBK/IPA (1:3), the exposed silicon was then etched by $\mathrm{SF}_6$ in reactive ion etcher (RIE) to define the bolometer pixel and prevent electrical shorting. Next, the antenna (5~nm Cr/25~nm Au) and contact electrodes (5~nm Cr/50~nm Au) were patterned by electron-beam lithography, followed by electron-beam evaporation and lift-off.
\vspace{1mm}

\noindent \textbf{LWIR photoresponse measurement.} The light from a 9.5 µm QCL (Alpes lasers) was guided into a Fourier-transform infrared spectrometer (FTIR) and then focused onto the device through a Cassegrain objective (15×, N.A. = 0.4) in a Hyperion 2000 microscope. The total power of the beam spot at different QCL driving currents was calibrated using both a thermoelectric detector (Thorlabs, S401C) placed under the Cassegrain objective and a liquid-cooled MCT detector, which collected the reflected power from the sample when the incident optical power was low. The incident beam was modulated either by a mechanical chopper at frequencies up to $3~\mathrm{kHz}$ or directly by the QCL driver at frequencies up to $400~\mathrm{kHz}$. The intensity profile of the beam spot was determined by scanning the SATE bolometer pixel with a step size of 2~μm over a $80\times80~\text{μm}^2$ region using the x-y motorized translation stage of the Hyperion 2000 microscope. The photocurrent of the device was probed by a custom-build probe station integrated with the motorized translation stage. A low-noise source meter (Keithley 2400) was used as a voltage source to define the operating point of the sample. The DC component was filtered from the photocurrent by an LC filter before entering the transimpedance amplifier (Stanford Research, SR570) to prevent saturation. A lock-in amplifier (Stanford Research, SR860), synchronized with the QCL driver, was used to obtain the peak-to-peak value of the photocurrent amplified by the TIA. Two neutral-density filters with attenuation factors of 10× and 1000×, together with tuning of the QCL driving current, were used to adjust the on-pixel incident power ($P_\mathrm{inc}$).
\vspace{1mm}

\noindent \textbf{Numerical simulations and calculation.}  The coupled optical-thermal simulation (detailed in Supplementary Section~2) was performed using COMSOL Multiphysics. The optical spectral response of the device was simulated using the Electromagnetic Waves, Frequency Domain (EWFD) interface under Gaussian-beam illumination with scattering boundary conditions. In addition to obtaining the reflection, extinction and absorption spectra of the device, the spatial distribution of the electromagnetic power-loss density was used as the heat-generation density to calculate the temperature-rise distribution within the pixel using the Heat Transfer in Solids (HTS) interface. The frequency response to dynamic heating was simulated in the same interface using a time-domain solver, where the heat-generation density was sinusoidally modulated in time. The solving time was chosen to allow the temperature oscillation to reach a harmonic steady state, from which the oscillation amplitude was extracted at different modulation frequencies to calculate the thermal frequency response. The breakdown threshold of the SATE bolometer pixel was calculated using the impact ionization integral (Supplementary Section~2.3). The empirical temperature- and field-dependent impact ionization coefficients for holes and electrons in silicon were used for the calculation of integral. Where, the lattice temperature was set to $300~\mathrm{K}$ and a one-dimensional device model with lateral  $p^+\text{-}p^{~}\text{-}n^+$ doping was used to calculate the width of depletion region and the electrical field distribution under external bias voltage.
\vspace{1 mm}

\noindent \textbf{Noise measurements.} The measurement of current noise in dark condition was performed using a probe station in a grounded electromagnetically shielded box (Everbeing). A low-noise source meter (Keithley 2400) was used as the voltage source to apply the voltage bias. Before connecting to the device, the voltage output from the source meter was connected to a second-order RC low-pass filter to suppress 60~Hz power-line noise and its harmonics. From the device side, this RC filter presents a low AC impedance, allowing the device noise current to pass through the bias branch with negligible loss. An LC filter was inserted between the device and the transimpedance amplifier (FEMTO, DHPCA-100) to separate the DC and AC components, with the noise current amplified by the TIA through the AC branch. An oscilloscope connected to the TIA output was used to record the time-domain current noise waveform, from which the noise spectrum was obtained using the Welch’s method (Supplementary Section~3.2). The noise measurement setup was validated by comparing the measured noise spectra density of carbon resistors with different resistances (2~$\mathrm{k}\Omega$, 5~$\mathrm{k}\Omega$, 10~$\mathrm{k}\Omega$, and 50~$\mathrm{k}\Omega$) to their theoretical Johnson current-noise spectral density, $\delta I_\mathrm{n,Johnson}=\sqrt{4k_\mathrm{B}T/R}$, showing good agreement up to 500~kHz.
\vspace{1.3mm}


\section*{Data Availability}
The data that support the plots within this paper and other findings of this study
are available from the corresponding author upon reasonable request.
\section*{Code Availability}
The computer code used to extrapolate the current noise amplitude spectral density of the sample in this paper is available from the corresponding author upon reasonable request.
\section*{Acknowledgments} This work is supported by the Office of Naval Research (Grant number: N000142412308), the Army Research Office (Grant number: W911NF-25-1-0238), the start-up grants from the CUNY Advanced Science Research Center and the CUNY Graduate Center. 

\section*{Authors Contributions}
Q.G. conceived the idea and supervised the project. W.D. designed and fabricated the devices and performed the measurements with assistance from G.H., Z.F., and Y.W. W.D. performed the simulations and device modeling. R.Y. and S.F. contribute to the data analysis and device modeling. W.D. and Q.G. wrote the manuscript with input from all authors.

\section*{Competing Interests}
The authors declare no competing interests.

\bibliographystyle{apsrev4-2}

\bibliography{references}

@book{howell2021thermal,
  title     = {Thermal Radiation Heat Transfer},
  author    = {Howell, John R. and Meng{\"u}{\c c}, M. Pinar and Daun, Kyle and Siegel, Robert},
  edition   = {7},
  publisher = {CRC Press},
  address   = {Boca Raton},
  year      = {2021},
   doi={10.1201/9780429327308},

}

@article{1033764,
  author={Tsuji, T. and Hattori, H. and Watanabe, M. and Nagaoka, N.},
  journal={IEEE Transactions on Intelligent Transportation Systems}, 
  title={Development of night-vision system}, 
  year={2002},
  volume={3},
  number={3},
  pages={203-209},
  doi={10.1109/TITS.2002.802927},
  publisher={IEEE}
}

@article{stewart2021nanophotonic,
  title={Nanophotonic engineering: a new paradigm for spectrally sensitive thermal photodetectors},
  author={Stewart, Jon W and Wilson, Nathaniel C and Mikkelsen, Maiken H},
  journal={ACS Photonics},
  volume={8},
  number={1},
  pages={71--84},
  year={2021},
  publisher={ACS Publications}
}

@article{watts2007microphotonic,
  title={Microphotonic thermal imaging},
  author={Watts, Michael R and Shaw, Michael J and Nielson, Gregory N},
  journal={Nature Photonics},
  volume={1},
  number={11},
  pages={632--634},
  year={2007},
   doi       = {10.1038/nphoton.2007.219},
  publisher={Nature Publishing Group UK London}
}

@article{govenius2016detection,
  title={Detection of zeptojoule microwave pulses using electrothermal feedback in proximity-induced Josephson junctions},
  author={Govenius, J and Lake, RE and Tan, KY and M{\"o}tt{\"o}nen, Mikko},
  journal={Physical Review Letters},
  volume={117},
  number={3},
  pages={030802},
  year={2016},
  doi       = {10.1103/PhysRevLett.117.030802},
  publisher={APS}
}

@article{jeon2021development,
  title={Development of a compact and robust mid-infrared spectrometer by using a silicon/air hyperspectral filter},
  author={Jeon, Taeyoon and Nateghi, Amirhossein and Jones, William Max and Choi, Changsoon and Cardenas, Juan Pablo and Ross, Charles and Scherer, Axel},
  journal={ACS Photonics},
  volume={9},
  number={1},
  pages={68--73},
  year={2021},
  doi       = {10.1021/acsphotonics.1c01750},
  publisher={ACS Publications}
}

@book{kruse2001uncooled,
  title={Uncooled thermal imaging: arrays, systems, and applications},
  author={Kruse, Paul W},
  volume={51},
  year={2001},
 doi       = {10.1117/3.415351},
  publisher={SPIE press}
}

@article{czerwinski2003activation,
  title={Activation energy analysis as a tool for extraction and investigation of p--n junction leakage current components},
  author={Czerwinski, A and Simoen, Eddy and Poyai, Amporn and Claeys, Corneel},
  journal={Journal of Applied Physics},
  volume={94},
  number={2},
  pages={1218--1221},
  year={2003},
  doi       = {10.1063/1.1598293},
  publisher={American Institute of Physics}
}

@article{abdel2019temperature,
  title={Temperature-dependent resistive properties of vanadium pentoxide/vanadium multi-layer thin films for microbolometer \& antenna-coupled microbolometer applications},
  author={Abdel-Rahman, Mohamed and Zia, Muhammad and Alduraibi, Mohammad},
  journal={Sensors},
  volume={19},
  number={6},
  pages={1320},
  year={2019},
  doi       = {10.3390/s19061320},
  publisher={MDPI}
}

@inproceedings{kohin2004performance,
  title={Performance limits of uncooled VOx microbolometer focal plane arrays},
  author={Kohin, Margaret and Butler, Neal R},
  booktitle={Infrared Technology and Applications XXX},
  volume={5406},
  pages={447--453},
  year={2004},
  doi       = {10.1117/12.542482},
  organization={SPIE}
}

@article{crowell1966temperature,
  title={Temperature dependence of avalanche multiplication in semiconductors},
  author={Crowell, CR and Sze, SM},
  journal={Applied Physics Letters},
  volume={9},
  number={6},
  pages={242--244},
  year={1966},
  doi       = {10.1063/1.1754731},
  publisher={AIP Publishing}
}

@inproceedings{niklaus2008mems,
  title     = {{MEMS}-based uncooled infrared bolometer arrays: a review},
  author    = {Niklaus, Frank and Vieider, Christian and Jakobsen, Henrik},
  booktitle = {{MEMS/MOEMS Technologies and Applications III}},
  series    = {Proceedings of SPIE},
  volume    = {6836},
  pages     = {68360D},
  publisher = {SPIE},
  year      = {2008},
  doi       = {10.1117/12.755128}
}

@article{rogalski2005hgcdte,
  author  = {Rogalski, A.},
  title   = {{HgCdTe} infrared detector material: history, status and outlook},
  journal = {Reports on Progress in Physics},
  volume  = {68},
  number  = {10},
  pages   = {2267--2336},
  year    = {2005},
  doi     = {10.1088/0034-4885/68/10/R01},
  publisher={APS}
}

@book{rogalski2000infrared,
  title     = {Infrared Detectors},
  author    = {Rogalski, Antoni},
  edition   = {1},
  publisher = {Gordon and Breach Science Publishers},
  address   = {Amsterdam},
  year      = {2000},
  doi       = {10.1201/9781420022506}
}

@article{richards2004bolometers,
    author = {Richards, P. L.},
    title = {Bolometers for infrared and millimeter waves},
    journal = {Journal of Applied Physics},
    volume = {76},
    number = {1},
    pages = {1-24},
    year = {1994},
    doi = {10.1063/1.357128},
}

@article{callen1951irreversibility,
  author  = {Callen, Herbert B. and Welton, Theodore A.},
  title   = {Irreversibility and Generalized Noise},
  journal = {Physical Review},
  volume  = {83},
  number  = {1},
  pages   = {34--40},
  year    = {1951},
  doi     = {10.1103/PhysRev.83.34},
  publisher  ={APS}
}

@article{mather1982bolometer,
  author  = {Mather, John C.},
  title   = {Bolometer noise: nonequilibrium theory},
  journal = {Applied Optics},
  volume  = {21},
  number  = {6},
  pages   = {1125--1129},
  year    = {1982},
  doi     = {10.1364/AO.21.001125}
}

@article{irwin1995application,
  author  = {Irwin, K. D.},
  title   = {An application of electrothermal feedback for high resolution cryogenic particle detection},
  journal = {Applied Physics Letters},
  volume  = {66},
  number  = {15},
  pages   = {1998--2000},
  year    = {1995},
  doi     = {10.1063/1.113674}
}

@article{Hooge1994,
  author={Hooge, F.N.},
  journal={IEEE Transactions on Electron Devices}, 
  title={1/f noise sources}, 
  year={1994},
  volume={41},
  number={11},
  pages={1926-1935},
  doi={10.1109/16.333808},
  publisher={IEEE}
}

@article{nyquist1928thermal,
  title={Thermal agitation of electric charge in conductors},
  author={Nyquist, Harry},
  journal={Physical Review},
  volume={32},
  number={1},
  pages={110},
  year={1928},
  publisher={APS},
  doi = {10.1103/PhysRev.32.110}
}

@article{Lee2022MotionBlurThermalDetector,
  author  = {Lee, Kangil and Ban, Yuseok and Kim, Changick},
  title   = {Motion Blur Kernel Rendering Using an Inertial Sensor: Interpreting the Mechanism of a Thermal Detector},
  journal = {Sensors},
  year    = {2022},
  volume  = {22},
  number  = {5},
  pages   = {1893},
  doi     = {10.3390/s22051893}
}

@article{chen2026bias,
  author  = {Chen, Jiazhen and Song, Yihao and Montealegre, David Alexander and Cai, Mingyang and Lee, Minjoo Larry and Xia, Fengnian},
  title   = {Bias-tunable temperature coefficient amplification beyond material limits in a single transistor},
  journal = {Nature Sensors},
  volume  = {1},
  pages   = {436--442},
  year    = {2026},
  doi     = {10.1038/s44460-026-00056-w}
}

@article{tan2010temperature,
  author  = {Tan, L. J. J. and Ong, D. S. G. and Ng, J. S. and Tan, C. H. and Jones, S. K. and Qian, Y. H. and David, J. P. R.},
  title   = {Temperature Dependence of Avalanche Breakdown in {InP} and {InAlAs}},
  journal = {IEEE Journal of Quantum Electronics},
  volume  = {46},
  number  = {8},
  pages   = {1153--1157},
  year    = {2010},
  doi     = {10.1109/JQE.2010.2044370}
}

@manual{teledyneflir2025bosonplus,
  author       = {{Teledyne FLIR}},
  title        = {{BOSON+}: High Performance, Uncooled, LWIR OEM Thermal Camera Module},
  organization = {Teledyne FLIR LLC},
  year         = {2025},
  url          = {https://www.flir.com/bosonplus}
}

@manual{lynred2022atto640d,
  author       = {{LYNRED}},
  title        = {{ATTO640D-02}: Infrared Detector Datasheet},
  organization = {LYNRED},
  year         = {2022},
  url          = {https://www.lynred.com/sites/default/files/2022-02/ATTO640D-02%20Datasheet.pdf}
}

@book{jaeger2002microelectronic,
  author    = {Jaeger, Richard C.},
  title     = {Introduction to Microelectronic Fabrication},
  edition   = {2},
  series    = {Modular Series on Solid State Devices},
  volume    = {5},
  publisher = {Prentice Hall},
  address   = {Upper Saddle River, NJ},
  year      = {2002},
  isbn      = {0-201-44494-1}
}

@article{grant1973ionization,
  author  = {Grant, W. N.},
  title   = {Electron and hole ionization rates in epitaxial silicon at high electric fields},
  journal = {Solid-State Electronics},
  volume  = {16},
  number  = {10},
  pages   = {1189--1203},
  year    = {1973},
  doi     = {10.1016/0038-1101(73)90147-0}
}

@article{kane1960zener,
  author  = {Kane, E. O.},
  title   = {Zener tunneling in semiconductors},
  journal = {Journal of Physics and Chemistry of Solids},
  volume  = {12},
  number  = {2},
  pages   = {181--188},
  year    = {1960},
  doi     = {10.1016/0022-3697(60)90035-4}
}

@article{vanderziel1955shotnoise,
  author  = {van der Ziel, A.},
  title   = {Theory of Shot Noise in Junction Diodes and Junction Transistors},
  journal = {Proceedings of the IRE},
  volume  = {43},
  number  = {11},
  pages   = {1639--1646},
  year    = {1955}
}

@article{mcintyre1966multiplication,
  author  = {McIntyre, R. J.},
  title   = {Multiplication noise in uniform avalanche diodes},
  journal = {IEEE Transactions on Electron Devices},
  volume  = {ED-13},
  number  = {1},
  pages   = {164--168},
  year    = {1966},
  doi     = {10.1109/T-ED.1966.15651}
}

@article{hooge1969oneoverf,
  author  = {Hooge, F. N.},
  title   = {1/f noise is no surface effect},
  journal = {Physics Letters A},
  volume  = {29},
  number  = {3},
  pages   = {139--140},
  year    = {1969},
  doi     = {10.1016/0375-9601(69)90076-0}
}

@article{chen2021ultrafast,
  author  = {Chen, Chen and Li, Cheng and Min, Seunghwan and Guo, Qiushi and Xia, Zhenyang and Liu, Dong and Ma, Zhenqiang and Xia, Fengnian},
  title   = {Ultrafast Silicon Nanomembrane Microbolometer for Long-Wavelength Infrared Light Detection},
  journal = {Nano Letters},
  volume  = {21},
  number  = {19},
  pages   = {8385--8392},
  year    = {2021},
  doi     = {10.1021/acs.nanolett.1c02972}
}

@book{ziman1979principles,
  title={Principles of the Theory of Solids},
  author={Ziman, John M},
  year={1979},
  doi       = {10.1017/CBO9781139644075},
  publisher={Cambridge university press}
}

@book{sze2021physics,
  title={Physics of semiconductor devices},
  author={Sze, Simon M and Li, Yiming and Ng, Kwok K},
  year={2021},
  url = {https://www.wiley.com/en-us/Physics+of+Semiconductor+Devices,+4th+Edition-p-9781119429111},
  publisher={John wiley \& sons}
}

@article{audhkhasi2025compact,
  title={Compact broadband thermal absorbers based on plasmonic fractal metasurfaces},
  author={Audhkhasi, Romil and Tara, Virat and Yu, Raymond and Povinelli, Michelle L and Majumdar, Arka},
  journal={Optics Letters},
  volume={50},
  number={18},
  pages={5750--5753},
  year={2025},
  doi     = {10.1364/OL.571991}

}

@article{xiao2021planck,
  title={Planck spectroscopy},
  author={Xiao, Yuzhe and Wan, Chenghao and Salman, Jad and Maywar, Ian J and King, Jonathan and Shahsafi, Alireza and Kats, Mikhail A},
  journal={Laser \& Photonics Reviews},
  volume={15},
  number={10},
  pages={2100121},
  year={2021},
  doi     = {10.1002/lpor.202100121},
  publisher={Wiley Online Library}
}

@inproceedings{hefner2001high,
  title={A high-speed thermal imaging system for semiconductor device analysis},
  author={Hefner, A and Berning, D and Blackburn, D and Chapuy, C and Bouche, Sebastien},
  booktitle={Seventeenth Annual IEEE Semiconductor Thermal Measurement and Management Symposium (Cat. No. 01CH37189)},
  pages={43--49},
  year={2001},
   doi     = {10.1109/STHERM.2001.915143},
  organization={IEEE}
}

@inproceedings{richards2005applications,
  title={Applications for high-speed infrared imaging},
  author={Richards, Austin A},
  booktitle={26th International Congress on High-Speed Photography and Photonics},
  volume={5580},
  pages={137--145},
  year={2005},
  organization={SPIE}
}

@inproceedings{skidmore2014uncooled,
  title={Uncooled microbolometers at DRS and elsewhere through 2013},
  author={Skidmore, George D and Han, CJ and Li, Chuan},
  booktitle={Image Sensing Technologies: Materials, Devices, Systems, and Applications},
  volume={9100},
  pages={910003},
  year={2014},
 doi     = {10.1117/12.2054135},
  organization={SPIE}
}

@article{shao2026nonlinear,
  title   = {Nonlinear exceptional points in an integrated acoustic-wave oscillator for longwave infrared sensing},
  author  = {Shao, Linbo and Xi, Zichen and Cen, Zengyu and Thomas, Joseph G. and Wang, Dongyao and Singh, Tanmay and Zhu, Liyan and Liu, Honghu and Ji, Jun and Yao, Yu and others},
  journal = {arXiv:2604.27371},
  year    = {2026},
  url     = {https://doi.org/10.48550/arXiv.2604.27371}
}

@article{hui2016plasmonic,
  title={Plasmonic piezoelectric nanomechanical resonator for spectrally selective infrared sensing},
  author={Hui, Yu and Gomez-Diaz, Juan Sebastian and Qian, Zhenyun and Alu, Andrea and Rinaldi, Matteo},
  journal={Nature Communications},
  volume={7},
  number={1},
  pages={11249},
  year={2016},
  doi     = {10.1038/ncomms11249},
  publisher={Nature Publishing Group UK London}
}

@article{yi2013plasmonically,
  title={Plasmonically enhanced thermomechanical detection of infrared radiation},
  author={Yi, Fei and Zhu, Hai and Reed, Jason C and Cubukcu, Ertugrul},
  journal={Nano Letters},
  volume={13},
  number={4},
  pages={1638--1643},
  year={2013},
   doi     = {10.1021/nl400087b},
  publisher={ACS Publications}
}

@article{xi2025room,
  title={Room-Temperature Mid-Infrared Detection Using Metasurface-Absorber-Integrated Phononic Crystal Oscillator},
  author={Xi, Zichen and Cen, Zengyu and Wang, Dongyao and Thomas, Joseph G and Srijanto, Bernadeta R and Kravchenko, Ivan I and Zuo, Jiawei and Liu, Honghu and Ji, Jun and Zhu, Yizheng and others},
  journal={Laser \& Photonics Reviews},
  volume={19},
  number={20},
  pages={e00498},
  year={2025},
  doi     = {10.1002/lpor.202500498},
  publisher={Wiley Online Library}
}

@article{vicarelli2022micromechanical,
  title={Micromechanical bolometers for subterahertz detection at room temperature},
  author={Vicarelli, Leonardo and Tredicucci, Alessandro and Pitanti, Alessandro},
  journal={ACS Photonics},
  volume={9},
  number={2},
  pages={360--367},
  year={2022},
  doi     = {10.1021/acsphotonics.1c01273},
  publisher={ACS Publications}
}

@article{das2023thermodynamically,
  title={Thermodynamically limited uncooled infrared detector using an ultra-low mass perforated subwavelength absorber},
  author={Das, Avijit and Mah, Merlin L and Hunt, John and Talghader, Joseph J},
  journal={Optica},
  volume={10},
  number={8},
  pages={1018--1028},
  year={2023},
  doi     = {10.1364/OPTICA.489761},
  publisher={Optica Publishing Group}
}

@article{xiao2020depth,
  title={Depth thermography: noninvasive 3D temperature profiling using infrared thermal emission},
  author={Xiao, Yuzhe and Wan, Chenghao and Shahsafi, Alireza and Salman, Jad and Kats, Mikhail A},
  journal={ACS Photonics},
  volume={7},
  number={4},
  pages={853--860},
  year={2020},
   doi      = {10.1021/acsphotonics.9b01588},
  publisher={ACS Publications}
}

@article{laurent201812,
  title={12-$\mu$ m-pitch electromechanical resonator for thermal sensing},
  author={Laurent, Ludovic and Yon, Jean-Jacques and Moulet, Jean-S{\'e}bastien and Roukes, Michael and Duraffourg, Laurent},
  journal={Physical Review Applied},
  volume={9},
  number={2},
  pages={024016},
  year={2018},
  publisher={APS},
  doi = {10.1103/PhysRevApplied.9.024016}
}

@article{rogalski2026new,
  title={New generation of uncooled thermal detectors: A review},
  author={Rogalski, Antoni},
  journal={Opto-Electronics Review},
  pages={e158924--e158924},
  year={2026},
  publisher={Polish Academy of Sciences (under the auspices of the Committee on~…},
  doi     = {10.24425/opelre.2026.158924}
}

@inproceedings{fisette2018customized,
  author    = {Fisette, Bruno and G{\'e}n{\'e}reux, Francis and B{\'e}land, David and Topart, Patrice and Tremblay, Mathieu and Desroches, Yan and Terroux, Marc and Marchese, Linda and Proulx, Christian and Picard, Francis and Dufour, Denis and Bergeron, Alain and Ch{\^a}teauneuf, Fran{\c{c}}ois and Alain, Christine},
  title     = {Customized packaged bolometers in niche applications at {INO}},
  booktitle = {Image Sensing Technologies: Materials, Devices, Systems, and Applications V},
  editor    = {Dhar, Nibir K. and Dutta, Achyut K.},
  volume    = {10656},
  pages     = {106560H},
  series    = {Proceedings of SPIE},
  year      = {2018},
  month     = may,
  publisher = {SPIE},
  doi       = {10.1117/12.2303513}
}

@article{bao2023heat,
  title={Heat-assisted detection and ranging},
  author={Bao, Fanglin and Wang, Xueji and Sureshbabu, Shree Hari and Sreekumar, Gautam and Yang, Liping and Aggarwal, Vaneet and Boddeti, Vishnu N and Jacob, Zubin},
  journal={Nature},
  volume={619},
  number={7971},
  pages={743--748},
  year={2023},
  doi     = {10.1038/s41586-023-06174-6},
  publisher={Nature Publishing Group UK London}
}

@article{brandao2001stability,
  title={Stability conditions, nonlinear dynamics, and thermal runaway in microbolometers},
  author={Brand{\~a}o, GB and De Almeida, LAL and Deep, Gurdip Singh and Lima, AMN and Neff, H},
  journal={Journal of Applied Physics},
  volume={90},
  number={4},
  pages={1999--2008},
  year={2001},
   doi     = {10.1063/1.1384852},
  publisher={American Institute of Physics}
}

@article{mousa2026ultrabroadband,
  title={Ultrabroadband Mid-to Long-Wave Infrared Spintronic Poisson Bolometer},
  author={Mousa, Mohamed A and Bauer, Leif and He, Daien and Gupta, Sakshi and Jape, Shubhankar and Singh, Utkarsh and Prasad, Bhagwati and Mukherjee, Partha P and Deka, Angshuman and Jacob, Zubin},
  journal={ACS Photonics},
  year={2026},
  url     = {https://pubs.acs.org/doi/abs/10.1021/acsphotonics.6c00028},
  publisher={ACS Publications}
}

@article{bauer2025exploiting,
  title={Exploiting spintronics at room temperature for long-wave infrared nanophotonic digital bolometers},
  author={Bauer, Leif and Deka, Angshuman and Mousa, Mohamed A and Gupta, Sakshi and He, Daien and Huang, Sijay and Prasad, Bhagwati and Santos, Tiffany and Ray, Biswajit and Jacob, Zubin},
  journal={Nano Letters},
  volume={25},
  number={14},
  pages={5599--5608},
  year={2025},
  doi     = {10.1021/acs.nanolett.4c05925},
  publisher={ACS Publications}
}

@article{yadav2022advancements,
  title={Advancements of uncooled infrared microbolometer materials: A review},
  author={Yadav, PV Karthik and Yadav, Isha and Ajitha, B and Rajasekar, Abraham and Gupta, Sudha and Reddy, Y Ashok Kumar},
  journal={Sensors and Actuators A: Physical},
  volume={342},
  pages={113611},
  year={2022},
  doi     = {10.1016/j.sna.2022.113611},
  publisher={Elsevier}
}

@article{gitelman2009cmos,
  title={CMOS-SOI-MEMS transistor for uncooled IR imaging},
  author={Gitelman, L and Stolyarova, S and Bar-Lev, S and Gutman, Z and Ochana, Y and Nemirovsky, Yael},
  journal={IEEE Transactions on electron devices},
  volume={56},
  number={9},
  pages={1935--1942},
  year={2009},
  doi     = {10.1109/TED.2009.2026523},
  publisher={IEEE}
}

@article{niklaus2008performance,
  title={Performance model for uncooled infrared bolometer arrays and performance predictions of bolometers operating at atmospheric pressure},
  author={Niklaus, Frank and Decharat, Adit and Jansson, Christer and Stemme, G{\"o}ran},
  journal={Infrared Physics \& Technology},
  volume={51},
  number={3},
  pages={168--177},
  year={2008},
  doi     = {10.1016/j.infrared.2007.08.001},
  publisher={Elsevier}
}

@article{hooge20021,
  title={1/f noise sources},
  author={Hooge, Friits N},
  journal={IEEE Transactions on Electron Devices},
  volume={41},
  number={11},
  pages={1926--1935},
  year={2002},
  doi     = {10.1109/16.333808},
  publisher={IEEE}
}

@article{huang2026thermal,
  title={Thermal detection of single photons using Dirac fermions},
  author={Huang, Bevin and Arnault, Ethan G and Jung, Woochan and Fried, Caleb and Russell, B Jordan and Watanabe, Kenji and Taniguchi, Takashi and Henriksen, Erik A and Englund, Dirk and Lee, Gil-Ho and others},
  journal={Nature Communications},
  year={2026},
  url     = {https://www.nature.com/articles/s41467-026-70648-0#citeas},
  publisher={Nature Publishing Group UK London}
}

\end{document}